\documentclass[12pt]{article}
\newcommand{\ba}{\begin{eqnarray}}
\newcommand{\ea}{\end{eqnarray}}
\newcommand{\nn}{\nonumber}
\def\lsim{\mathrel{\raise.3ex\hbox{$<$\kern-.75em\lower1ex\hbox{$\sim$}}}}
\def\gsim{\mathrel{\raise.3ex\hbox{$>$\kern-.75em\lower1ex\hbox{$\sim$}}}}

\newcommand{\op}[1]{\left(r\,\frac{d}{dr}+#1\right)}
\newcommand{\mors}[0]{\left(\frac{M}{r^2}\right)}
\newcommand{\oh}[0]{\frac{1}{2}}
\newcommand{\we}[4]{e^{#1}\wedge e^{#2} \wedge e^{#3}\wedge e^{#4}}
\newcommand{\rr}[1]{R^{#1}_{#1}}
\begin{document}
\title{Non-factorisable metrics and Gauss--Bonnet terms in higher
dimensions}
\author{
{\large B. Abdesselam}$^{1)}$\thanks{e-mail address:
boucif@yahoo.fr}, {\large A. Chakrabarti}$^{2)}$\thanks{e-mail
address: Amitabha.Chakrabarti@cpht.polytechnique.fr}, {\large J.
Rizos}$^{3)}$\thanks{e-mail address: irizos@cc.uoi.gr}, and
{\large D. H. Tchrakian}$^{4)}$\thanks{e-mail address:
tigran@thphys.may.ie} \\ \\
$^{1)}${\small Laboratoire de Physique Th\'eorique`,
Centre Universitaire Moustapha Stambouli,} \\
{\small 29000-Mascara, Alg\'erie}\\ \\
$^{2)}${\small Centre de Physique Th\'eorique, Ecole
Polytechnique, F-91128 Palaiseau Cedex, France}\\ \\
$^{3)}${\small Department of Physics, University of Ioannina,
GR-45110 Ioannina, Greece} \\ \\
$^{4)}${\small Department of
Mathematical Physics, National University of Ireland Maynooth,} \\
{\small Maynooth, Ireland} \\
{\small and} \\
{\small School of Theoretical Physics -- DIAS, 10 Burlington
Road, Dublin 4, Ireland}}

\maketitle
\begin{abstract}
An iterative construction of higher order Einstein tensors for a
maximally Gauss-Bonnet extended gravitational Lagrangian was
introduced in a previous paper. Here the formalism is extended to
non-factorisable metrics in arbitrary ($d+1$) dimensions in the
presence of superposed Gauss-Bonnet terms. Such a generalisation
turns out to be remarkably convenient and elegant. Having thus
obtained the variational equations we first construct bulk
solutions, with nonzero and zero cosmological constant. It is also
pointed out that in the absence of Gauss-Bonnet terms a
Schwarzschild type solution can be obtained in the
non-factorisable case. Two positive tension branes are then
inserted and their tensions are obtained in terms of parameters in
the warp factor. Relations to recent studies of several authors
are pointed out.
\end{abstract}
\newpage
 {\setcounter{equation}{0}}
\section{Introduction}
A particularly helpful formulation was presented in
\cite{Chakrabarti:2001di} for
explicit construction of static spherically symmetric metrics
(generalizing Schwarzschild and de-Sitter type solutions) in
$(d-1)$ space dimensions in presence of higher derivative
Gauss--Bonnet (GB) terms in the gravitational Lagrangian. The key
feature was an iterative construction of the Riemann, Ricci and
Einstein tensors to incorporate the contributions of GB terms of
successive orders. This provides a powerful and elegant approach
systematizing the combinatorics involved when the maximal number
of GB terms are present. The solutions can be displayed
systematically.

In four dimensional theories the GB term is topological, namely a total
divergence, and hence contributes nothing to
the classical equations of motion. However,
in the context of  higher dimensional theories the GB invariants
are non-trivial and their presence  is required for obtaining a
ghost-free theory of gravity. For example, in  string theory, the
GB combination appears naturally in the tree-level effective
action of  heterotic superstring at the next-to-leading order in the
$\alpha'$ (string-tension) expansion \cite{effstr}.

In the last years, considerable effort   has been devoted to the
study of higher dimensional space-times in the presence of
non-factorizable metrics \cite{nf,Kogan:2000vb,R}.
The effect of the  GB invariant has also been considered in the recent
literature \cite{allgb,loc,gbcosmo}.
In this article, we  study the consequences of  the presence of
maximal GB terms in the context of
Randall--Sundrum type metrics in higher dimensions, utilizing the
techniques developed in \cite{Chakrabarti:2001di}. More
specifically, we show how the bigravity model with positive
tension branes of Ref. \cite{Kogan:2000vb} generalises to
arbitrary dimensions in the presence of maximal number of GB
terms.

In Sec. {\bf 2} we start with the following metric in $d+1$ dimensions
\ba
ds^2=f(y)\left\{\mp(1-L(r))\,dt^2+\frac{1}{(1-L(r))}\,dr^2+r^2\,
d\Omega^2_{d-2}\right\}
+dy^2\label{1.1}
\ea
where $f(y)$ and $L(r)$ are, to start with, unknown functions. We show
how the formalism of \cite{Chakrabarti:2001di} smoothly generalises to
yield the Einstein tensors in the presence of the maximal number of GB
terms superposed to give the generalised Lagrangian. Then in Sec. {\bf 3}
bulk solutions are presented with
\ba
L(r)&=&br^2 \nonumber \\
f(y)&=&c_0+c_1e^{2ky}+c_2e^{-2ky}\ , \label{1.2}
\ea
the constant parameters satisfying
\ba
c_0^2=4c_1c_2\quad ,\quad b=-2c_0k^2\label{1.3}\ .
\ea

Now, with (\ref{1.2}) and (\ref{1.3}), (\ref{1.1}) solves the generalised
gravitational equation
\ba
G^a{}_b=\sum_{p=1}^{P}\kappa_{(p)}G_{(p)}{}^a{}_b=\Lambda\ \delta^a{}_b\ ,
\quad 2P\le d+1\ ,\label{1.4}
\ea
where $\kappa_{(p)}$ is the coupling strength of the $2p$-th order Ricci
scalar in the Lagrangian
\ba
{\cal L}\ =\ \sum_{p=1}^P\frac{1}{2p} \sqrt{\mp g}\ \kappa_{(p)}\ R_{(p)}
\ea
and $\Lambda$ is the bulk cosmological constant. Eqn. (\ref{1.4}) is
satisfied provided that $(k^2)$ satisfies the polynomial equation
\ba
\frac{1}{2}\kappa_{(1)}\,d(d-1)(k^2)-
\frac{1}{8}\kappa_{(2)}\,d\,(d-1)\,(d-2)\,
(d-3)\,(k^2)^2+&~&\nn\\
\dots+(-1)^{P+1}\frac{1}{2^P\,P!}\,d\,(d-1)\dots+
(d-2P+1)(k^2)^P&=&\Lambda\label{1.6}
\ea
{\it Thus, as more GB terms are superposed (with more nonzero
$\kappa_{(P)}$) the form of the metric} (\ref{1.1}) {\it is conserved but
$(k^2)$ has to satisfy a correspondingly higher order polynomial
equation} (\ref{1.6}).

This crucial constraint on the warp factor $f(y)$ (\ref{1.2}) is extracted
in a remarkably compact and convenient fashion in our formalism. Further
comments on (\ref{1.6}) can be found in Sec. {\bf 2}.

We consider both cases
\ba
\Lambda\ \neq \ 0\quad {\rm and}\quad \Lambda\ =0\ .\label{1.7}
\ea

We also point out a possibility usually ignored. For $P=1$ (i.e. {\it
in the absence of GB terms}) one can generalise $L(r)$ in (\ref{1.2}) to
\ba
L(r)=\frac{c}{r^{d-3}}+br^2\ .\label{1.8}
\ea
{\it One thus obtains a Schwarzschild type black hole in the
non-factorisable case}.

In Sec. {\bf 4} we insert the $(d-1)$-branes by changing, to start with,
$y$ to $|y|$ in (\ref{1.2}) and writing $f(y)$
\ba
f(y)=(a(y))^2=\left(\frac{\cosh k(y_0-|y|)}{\cosh ky_0}\right)^2\ .
\label{1.9}
\ea
This corresponds, along with $|y|$ for $y$, to
\ba
(c_0\ ,c_1\ ,c_2)=(e^{ky_0}+e^{-ky_0})^{-2}\
\left(2\ ,e^{-2ky_0}\ ,e^{2ky_0}\right)\ ,\label{1.10}
\ea
satisfying the constraint
\[
c_0^2\ =\ 4c_1c_2\ ,
\]
in (\ref{1.3}). The close relation to Ref. \cite{Kogan:2000vb} is now
evident, and the brane tensions (of the two positive tension branes at
$y=0$ and at $y=Y$, say) are obtained in a similar fashion.

It is shown, somewhat analogously to (\ref{1.6}), that the effect of the
GB terms in the resulting brane tensions appears in the polynomial factors
\ba
K_{(P)}=\sum_{p=1}^P\kappa_{(p)}\frac{(-1)^{p-1}}{2^{p-1}\,(p-1)!}\,(d-1)
\dots(d-2p+1)\ (k^2)^{p-1}
\ea
in the delta function terms. In our formalism this crucial feature is also
remarkably easily obtained. We defer other references and comments to the
concluding section.

{\setcounter{equation}{0}}
\section{Ansatz for a nonfactorizable metric and construction of
the corresponding GB terms}

We start with a metric containing two
unknown functions  (to be indicated below) and construct
explicitly and systematically the Riemann, Ricci and Einstein
tensor and their GB generalizations (involving the above-mentioned
functions and their derivatives in specific fashions). They will
serve as inputs for the bulk Lagrangian to be defined in the
following section leading to exact bulk solutions of the
variational equations. Consequences of insertions of branes will
be considered next.

The coordinates will be denoted as
\ba
\left(t,r,\theta_1,\theta_2,\dots,\theta_{d-2},y\right)
\ea
where $\left(\theta_1,\theta_2,\dots,\theta_{d-2}\right)$  are the
angular coordinates. The dimension is $D=d+1$.

Let
\ba
ds^2=f(y)\left\{\mp(1-L(r))\,dt^2+\frac{1}{(1-L(r))}\,dr^2+r^2\,
d\Omega^2_{d-2}\right\}
+dy^2
\ea
where
\ba
\Omega^2_{d-2}=\sum_{i=1}^{d-2}P_i^2\,d\theta_i
\ea
and $P_1=1, P_i=\prod_{k=1}^{i-1}\sin\theta_k, i=2,\dots,d-2$.
Choices for $f(y)$ and $L(r)$ will be specified in the next
section. We will usually suppress the arguments $y$ and $r$ in
solutions below. Tangent plane vectors (vielbeins) are the
$1$-forms
\ba
e^t=\sqrt{f(1-L)}\,dt,\ e^r=\sqrt{\frac{f}{1-L}}\,dr,\
e^i=\sqrt{f}\,r\,P_i\,d\theta_i,\ e^y=dy,\ i=1,\dots,d-2\label{vb}
\ea
Tangent plane indices $(a,b,\dots)$ will be continued to be
denoted by $(t,r,i_1,\dots,i_{d-2},y)$ instead of, say, by
$(\hat{t},\hat{r},\dots)$. For our diagonal metric there will be
no confusion. The indices $(a,b,\dots)$ namely $(t,r,\dots)$ as
tangent plane ones will be raised and lowered by $\eta_a^b$ rather than
$g^{\mu}_{\nu}$.

The metric being diagonal, the spin connection $1-$forms are
\ba
\omega^{ab}=-\omega^{ba}=\frac{1}{\sqrt{g_{aa}\,g_{bb}}}
\left\{\left(\partial_b\sqrt{g_{aa}}\right)-
\left(\partial_a\sqrt{g_{bb}}\right)\right\}, \ \ a,b \ \ \mbox{not
summed}\label{sc}
\ea
the Riemann tensor $2-$forms are
\ba
R^{ab}=d\omega^{ab}+\omega^{ac}\wedge\omega_c^b\label{rm}
\ea
Define,
\ba
h_1(y)&=&\frac{1}{4f^2}\left(\frac{df}{dy}\right)^2\label{hone}\\
h_2(y)&=&\frac{1}{4f^2}\left(2f\,\frac{d^2f}{dy^2}-
\left(\frac{df}{dy}\right)^2\right)\label{htwo}
\ea
and
\ba
M(r,y)=\frac{1}{f(y)}\left(L(r)-r^2\,f(y)\,h_1(y)\right)\label{mm}
\ea
Suppressing arguments will usually write
\ba
M=\frac{1}{f}\left(L-r^2\,f\,h_1\right)
\ea
and
$M'=\frac{\partial M(r,y)}{\partial r}$, $M''=\frac{\partial^2
M(r,y)}{\partial r^2}$.

From (\ref{vb}), (\ref{sc}) and  (\ref{rm}) one obtains the
nonvanishing components of (\ref{rm}) (with $i=1,\dots,d-2$) as
\ba
R^{tr}&=&\frac{1}{2}\,M'' e^t\wedge e^r
=\left\{\frac{1}{2}\left(r\frac{d}{dr}+2\right)\left(r\frac{d}{dr}
+1\right)\left(\frac{M}{r^2}\right)\right\}e^t\wedge e^r\nn\\
R^{ti}&=&\frac{1}{2r}\,M' e^t\wedge
e^i=\frac{1}{2}\left(r\frac{d}{dr}+2\right)\left(\frac{M}{r^2}\right)
e^t\wedge e^i\nn\\
R^{ri}&=&\frac{1}{2}\left(r\frac{d}{dr}+2\right)\left(\frac{M}{r^2}\right)
e^r\wedge e^i\label{rcs}\\
R^{ij}&=&\left(\frac{M}{r^2}\right)\,e^i\wedge e^j\nn\\
R^{ty}&=&-h_2\,e^t\wedge e^y\nn\\
R^{ry}&=&-h_2\,e^r\wedge e^y\nn\\
R^{iy}&=&-h_2\,e^i\wedge e^y\nn
\ea
As compared to \cite{Chakrabarti:2001di}, in $(R^{tr}$, $R^{ti}$, $R^{ri})$ $M$
replaces $L$ and there are the extra components $(R^{ty}$, $
R^{ry}$, $R^{iy})$ each proportional to $h_2$. Each $R^{ab}$ still
has only one ``diagonal" component $R^{ab}_{ab}$. Hence the
Ricci-tensor components are easily obtained as follows. The
nonzero (all diagonal) terms (adding a subscript (1) in view of
subsequent generalization to higher order GB terms) are
\footnote{Note that for $L=c\,r^2$, $f=c\,y^2$, one obtains $M=0=h_2$
and hence a flat space.}
\ba
{R_{(1)}}^t_t&=&\sum_a
R_{ta}^{ta}=R^{tr}_{tr}+\sum_{i=1}^{d-2}R_{ti}^{ti}+R_{ty}^{ty}=
\frac{1}{2}\left(r\frac{d}{dr}+2\right)\left(r\frac{d}{dr}+(d-1)\right)
\left(\frac{M}{r^2}\right)-h_2\nn\\
{R_{(1)}}^r_r&=&{R_{(1)}}^t_t\ref{rcs}\\
{R_{(1)}}^i_i&=&R^{ti}_{ti}+R^{ri}_{ri}+\sum_{j\ne
i}R^{ij}_{ij}+R_{iy}^{iy}=\left(r\frac{d}{dr}+(d-1)\right)
\left(\frac{M}{r^2}\right)-h_2\nn\\
{R_{(1)}}^y_y&=&-d\,h_2\nn
\ea
The Ricci scalar is
\ba
R_{(1)}={R_{(1)}}_t^t+{R_{(1)}}_r^r+\sum_{i=1}^{d-2}{R_{(1)}}_i^i+{R_{(1)}}_y^y=
\left(r\frac{d}{dr}+d\right)\left(r\frac{d}{dr}+(d-1)\right)
\left(\frac{M}{r^2}\right)-2\,d\,h_2
\ea
The nonzero components of the corresponding Einstein tensor
\ba
{G_{(1)}}^a_b={R_{(1)}}^a_b-\frac{1}{2}\,\eta^a_b\,R_{(1)}\nn
\ea
are obtained as
\ba
{G_{(1)}}^t_t&=&-\frac{1}{2}(d-2)\op{(d-1)}\mors+(d-1)\,h_2\nn\\
{G_{(1)}}^r_r&=&{G_{(1)}}^t_t\nn\\
{G_{(1)}}^i_i&=&-\oh\,\op{(d-2)}\,\op{(d-1)}\,\mors+(d-1)\,h_2\label{gcs}\\
{G_{(1)}}^y_y&=&-\oh\,\op{d}\,\op{(d-1)}\mors\nn
\ea
Now we show how the recursion formula used in \cite{Chakrabarti:2001di}
simplifies here also the construction of GB terms of order $2 p$
$(p=2,\dots,P, 2P\le D)$. (For $p=1$ one has the standard $R^{ab}$
obtained above.) Totally antisymmetrized $2p$-forms are obtained
as follows. We continue to use exclusively tangent plane indices.

For $p=2$,
\ba
R^{abcd}=R^{ab}\wedge R^{cd}+R^{ad}\wedge R^{bc} +R^{ac}\wedge
R^{db}\label{rmd}
\ea
the indices $(b,c,d)$ being circularly permuted. For arbitrary $p$
\ba
R^{a_1 a_2 \dots a_{2p}}=R^{a_1 a_2}\wedge R^{a_3\dots a_{2p}}+
\left(\mbox{circ. perm. of } a_2,\dots,a_{2p}\right)\label{rcp}
\ea
giving $(1\cdot3\cdot5\cdot\dots\cdot(2p-1))$ terms of the type
\ba
R^{a_1 a_2}\wedge R^{a_3 a_4}\wedge\dots\wedge R^{a_{2p-1}
a_{2p}}\nn\ .
\ea
For $p=2$, implementing (\ref{rcs}) in (\ref{rmd}),
\ba
R^{trij}&=&R^{tr}\wedge R^{ij}+r^{tj}\wedge R^{ri}+ R^{ti}\wedge
R^{jr}\nn\\
&=&\frac{1}{4}\op{4}\op{3}\mors^2\,\we{t}{r}{i}{j}\ .
\ea
Proceeding similarly for the other members of $R^{abcd}$, the
nonzero (all diagonal) components are obtained as
\ba
R^{trij}_{trij}&=&\frac{1}{4}\,\op{4}\op{3}\mors^2\nn\\
R^{tijk}_{tijk}&=&R^{rijk}_{rijk}=\frac{3}{4}\,\op{4}\mors^2\nn\\
R^{ijkl}_{ijkl}&=&3\mors^2 \label{rmcs}\\
R^{triy}_{triy}&=&-\oh\, h_2\op{2} \op{3} \mors\nn\\
R^{tijy}_{tijy}&=&R^{rijy}_{rijy}=-h_2\,\op{3}\mors\nn\\
R^{ijky}_{ijky}&=&-3\,h_2\mors\nn
\ea
(Simultaneous permutations of the same top and bottom indices
leave the values unchanged.)

Now the nonzero components of the $p=2$ Ricci tensor components are
obtained as follows\footnote{In the last step there is no sum over
$i,j,\dots$. The values of $R^{trij}_{trij}$ for different $(ij)$ being
the same $\sum_{i,j}$ is replaced by $\left(d-2\atop 2\right)$.
Similarly for other terms.}
\ba
{R_{(2)}}^t_t&=&\sum_{i,j}R^{trij}_{trij}+\sum_{i,j,k}
R^{tijk}_{tijk}+\sum_{i}R^{triy}_{triy}+\sum_{i,j}
R^{tijy}_{tijy}\nn\\
&=&\left(d-2\atop 2\right) R^{trij}_{trij}+\left(d-2\atop 3\right)
R^{tijk}_{tijk}+(d-2) R^{triy}_{triy}+\left(d-2\atop 2\right)
R^{tijy}_{tijy}
\ea
Thus
\ba
{R_{(2)}}^t_t &=&
\frac{1}{8}\,(d-2)(d-3)\op{4}\op{(d-1)}\mors^2\nn\\
&-&\oh\,h_2(d-2)\op{3}\op{(d-1)}\mors
\ea
Proceeding similarly
\ba
{R_{(2)}}^r_r &=&{R_{(2)}}^t_t
\ea
and
\ba
{R_{(2)}}^i_i &=&\sum_{j}
R^{itrj}_{itrj}+\sum_{j,k}\left(R^{itjk}_{itjk}+R^{irjk}_{irjk}\right)+
\sum_{j,k,l}R^{ijkl}_{ijkl}+R^{itry}_{itry}\nn\\
&~&{}+\sum_j\left(R^{itjy}_{itjy}+
R^{irjy}_{irjy}\right)+\sum_{j,k}R^{ijky}_{ijky}
\ea
Here, $i$ being fixed, the sums over $(j)$, $(j,k)$ and $(j,k,l)$
are replaced respectively by $\left(d-3\atop1\right)$,
$\left(d-3\atop2\right)$ and $\left(d-3\atop3\right)$.

Finally,
\ba
{R_{(2)}}^i_i&=&\frac{1}{4}\,(d-3)\op{2(d-2)}\op{(d-1)}\mors^2\nn\\
&~&-\oh\,h_2\op{3(d-2)}\op{(d-1)}\mors
\ea
Similarly,
\ba
{R_{(2)}}^y_y&=&\sum_i
\rr{ytri}+\sum_{i,j}\left(\rr{ytij}+\rr{yrij}\right)
+\sum_{i,j,k}\rr{yijk}\nn\\
&=&-\oh(d-2)h_2\op{d}\op{(d-1)}\mors
\ea
The $p=2$ Ricci scalar is
\ba
R_{(2)}&=&{R_{(2)}}^t_t+{R_{(2)}}^r_r+\sum_i
{R_{(2)}}^i_i+{R_{(2)}}^y_y
\nn\\
&=&\oh(d-2)(d-3)\op{d}\op{(d-1)}\mors^2\nn\\
&~&-2\,h_2(d-2)\op{d}\op{(d-1)}\mors
\ea
The Einstein tensor for arbitrary $p$ is (since $\eta^a_b=1$ for
our indices)
\ba
{G_{(p)}}^a_b={R_{(p)}}^a_b-\frac{1}{2p}R_{(p)}\label{et}
\ea
Hence for $p=2$,
\ba
{G_{(2)}}^t_t&=&{R_{(2)}}^t_t-\frac{1}{4}R_{(2)}\ (={G_{(2)}}^r_r)\nn\\
&=&-\frac{1}{8}\,(d-2)(d-3)(d-4)\op{(d-1)}\mors^2\nn\\
&~&+\oh\,h_2\,(d-2)(d-3)\op{(d-1)}\mors\label{2.27}\\
{G_{(2)}}^i_i&=&{R_{(2)}}^i_i-\frac{1}{4}R_{(2)}\nn\\
&=&-\frac{1}{8}\,(d-3)(d-4)\op{(d-2)}\op{(d-1)}\mors^2\nn\\
&~&+\oh\,h_2\,(d-3)\op{(d-2)}\op{(d-1)}\mors\label{2.28}\\
{G_{(2)}}^y_y&=&{R_{(2)}}^y_y-\frac{1}{4}R_{(2)}\nn\\
&=&-\frac{1}{8}\,(d-2)(d-3)\op{d}\op{(d-1)}\mors^2\label{2.29}
\ea
Successive iterations, using (\ref{rcp}), can be shown to lead for
arbitrary $p$ $(2p\le d)$ to the nonzero components
${R_{(p)}}^a_b$ given below. For $p>2$, one obtains
\ba
{R_{(p)}}^t_t&=&{R_{(p)}}^r_r=\nn\\
&=&\frac{1}{2^p\,p!}(d-2)(d-3)\dots(d-2p+1)\op{2p}\op{(d-1)}\mors^p\nn\\
&~&-\frac{1}{2^{p-1}(p-1)!}h_2(d-2)\dots(d-2p+2)\nn\\
&~&\times\op{(2p-1)}\op{(d-1)}\mors^{p-1}
\ea
\ba
{R_{(p)}}^i_i&=&\frac{1}{2^p\,p!}(d-3)\dots(d-2p+1)
\left((p-1)r\,\frac{d}{dr}+p(d-2)\right)\op{(d-1)}\mors^p\nn\\
&~&-\frac{1}{2^{p-1}(p-1)!}\,h_2\,(d-3)\dots(d-2p+2)\nn\\
&~&\times\left((2p-3)\frac{d}{dr}+
(2p-1)(d-2)\right)\op{(d-1)}\mors^{p-1}
\ea
\ba
{R_{(p)}}^y_y&=&-\frac{1}{2^{p-1}(p-1)!}\,h_2\,(d-2)\dots(d-2p+2)
\op{d}\op{(d-1)}\mors^{p-1}
\ea
Hence,
\ba
{R_{(p)}}&=&{R_{(p)}}^t_t+{R_{(p)}}^r_r
+\sum_i {R_{(p)}}^i_i+{R_{(p)}}^y_y\nn\\
&=&\frac{1}{2^{p-1}(p-1)!}(d-2)\dots(d-2p+1)\op{d}\op{(d-1)}\mors^p\nn\\
&~&-\frac{1}{2^{p-1}(p-1)!}\,h_2\,(d-2)\dots(d-2p+2)\op{d}\op{(d-1)}\mors^{p-1}
\ea
Now using (\ref{et}), namely
\[
{G_{(p)}}^a_b={R_{(p)}}^a_b-\frac{1}{2p}R_{(p)}\ ,
\]
\ba
{G_{(p)}}^t_t&=&{G_{(p)}}^r_r\nn\\
&=&-\frac{1}{2^p\,p!}(d-2)\dots(d-2p+1)\nn\\
&~&\times\op{(d-1)}\left\{(d-2p)\mors-2\,p\,h_2\right\}
\mors^{p-1}\label{gptt}\\
{G_{(p)}}^i_i&=&-\frac{1}{2^p\,p!}(d-3)\dots(d-2p+1)\nn\\
&~&\times\op{(d-2)}\op{(d-1)}\left\{(d-2p)\mors-2\,p\,h_2\right\}
\mors^{p-1}\label{gpii}\\
{G_{(p)}}^y_y&=&-\frac{1}{2^p\,p!}(d-2)\dots(d-2p+1)\times\op{d}\op{(d-1)}\mors^p
\label{gpyy}
\ea
These results should be compared with the corresponding results in
\cite{Chakrabarti:2001di} (Eqns (4.1),(4.2),(4,3) of
\cite{Chakrabarti:2001di} give the summed
up contribution of all $G_{(p)}$). \setcounter{equation}{0}

{\setcounter{equation}{0}}
\section{The bulk Lagrangian and solutions}
With suitable
choices of units and sign conventions, the bulk Lagrangian with a
maximal number of GB terms is defined to be
\ba
{\cal L}=\sum_{p=1}^P\frac{1}{2p}\sqrt{\mp g}\ \kappa_{(p)}\,R_{(p)}\ ,
\ \ \  (2P\le D)
\ea
(For odd $d$, the GB term with $2P=D=d+1$, does not contribute to
the equations of motion but has a topological significance. For
$2p>(d+1)$, $R^{a_1 a_2\dots a_{2p}}$ and hence $R_{(p)}$ is
identically zero due to total antisymmetry.)

The constants $\kappa_{(p)}$ are arbitrary, but we will usually
assume
\ba
\kappa_{(1)}\gg \kappa_{(p)}\ \ (p>1)\label{int}
\ea
and possibly
\ba
\kappa_{(1)}\gg \kappa_{(2)} \gg \kappa_{(3)}\gg\dots\label{inm}
\ea
and so on. The variational equations (in the absence of sources
and matter fields)
\ba
G^a_b=\sum_{p=1}^P\kappa_{(p)}\,{G_{(p)}}^a_b=\Lambda\,
\delta^a_b\label{gkl}\label{3.4}
\ea
(with tangent plane indices) where $\Lambda$ is some bulk
cosmological constant in suitable units.

\subsection{Bulk solutions for superposed GB terms}
It was shown in \cite{Chakrabarti:2001di} how to take into account the
effect of a term $\sim r^2$ in $L(r)$ (see Sec. VII of
\cite{Chakrabarti:2001di}). Here we show, if $L(r)$ has only this
AdS (or dS) type term, how $f(y)$ can be chosen to provide exact
solutions.

Set
\ba
L= b\,r^2\label{lbr}
\ea
\ba
f=c_0+c_1\,e^{2 k y}+c_2 e^{-2 k y}\label{fcc}
\ea
Then from (\ref{hone}), (\ref{htwo}) and (\ref{mm})
\ba
h_1&=&k^2\left(1-\frac{2 c_0}{f}+\frac{c^2-4 c_1 c_2}{f^2}\right)\nn\\
h_2&=&k^2\left(1+\frac{c^2-4 c_1 c_2}{f^2}\right)\label{fccs}\\
\mors&=&-k^2+\frac{b+2 c_0\,k^2}{f}-k^2\frac{c^2-4 c_1
c_2}{f^2}\nn
\ea
Hence for
\ba
c_0^2=4\,c_1\,c_2\ ,\qquad b=-2\,c_0\,k^2\label{cc}
\ea
one obtains, along with $(1-L(r))=1+2c_0k^2r^2$ in the metric
\ba
h_2=-\frac{M}{r^2}=k^2\ .\label{tn}
\ea
(Note that (\ref{cc}) is invariant under the exchange of $c_1$,
$c_2$, namely for $k\to -k$ in (\ref{lbr}). The equations below
will involve $k^2$.

Now, from (\ref{gptt}), (\ref{gpii}) and
(\ref{gpyy}), the nonzero components of ${G_{(p)}}^a_b$ all
coincide yielding
\ba
{G_{(p)}}^t_t={G_{(p)}}^r_r={G_{(p)}}^y_y=(-1)^{(p+1)}\frac{1}{2^p\,p!}d\,(d-1)\dots
(d-2p+1)(k^2)^p\ ,\label{tten}
\ea
whence (\ref{gkl}) reduces to
\ba
\frac{1}{2}\kappa_{(1)}\,d(d-1)(k^2)-
\frac{1}{8}\kappa_{(2)}\,d\,(d-1)\,(d-2)\,
(d-3)\,(k^2)^2+&~&\nn\\
\dots+(-1)^{P+1}\frac{\kappa_{(P)}}{2^p\,p!}\,d\,(d-1)\dots+
(d-2 P+1)(k^2)^P&=&\Lambda\ .\label{tel}
\ea
{\it Thus it seen that in the context of} (\ref{lbr}), (\ref{fcc}) {\it
and} (\ref{cc}), {\it the effect of the BG terms reduce to the fact that
$k^2$ is constrained to satisfy a polynomial equation of degree $P$.}
Substituting this $k^2$ in (\ref{cc}), for $c_0>0$ (i.e.
$(1-L)=1+2c_0k^2r^2$), one has an AdS type metric.

For $P=1$, one has (in the absence of GB terms),
\ba
k^2=\frac{2\Lambda}{\kappa_{(1)}d(d-1)}\label{kl}
\ea
For $P=2$, one obtains
\ba
\frac{1}{2}\kappa_{(1)}d(d-1)(k^2)-
\frac{1}{8}\kappa_{(2)}d(d-1)(d-3)(k^2)^2=\Lambda
\label{tf}
\ea
or
\ba
(k^2)=\frac{2\kappa_{(1)}}{\kappa_{(2)}(d-2)(d-3)}\left\{1-\left(
1-2\Lambda\frac{\kappa_{(2)}(d-2)(d-3)}{\kappa_{(1)}^2 d (d-1)}
\right)^{\frac{1}{2}}\right\}
\ea
The sign before the square root has been chosen to obtain
(\ref{kl}) as the leading term on expanding the square root in
powers of $\frac{\kappa_{(2)}}{\kappa_{(1)}}$.

For $P>2$, if (\ref{int}) and (\ref{inm}) are maintained, one has to
select a solution satisfying similarly consistency (developing in powers
of $\kappa_{(2)},\kappa_{(2)}^2, \kappa_{(2)}\kappa_{(3)},\dots$) with
the lower degree cases.

For $d=10$, keeping all possible $p$'s ($p=1,2,3,4,5$) one obtains $k^2$
in terms of {\em elliptic functions}. For $d>10$, {\em hyperelliptic
functions} are needed. (Compare the discussion of Section V of
\cite{Chakrabarti:2001di}, where such functions arise for $(L/r^2)$ and
the horizon $r_H$.)

\subsection{Special cases}
(1) \underline {$\Lambda=0$}

For zero bulk cosmological constant and for $P>1$ one still obtains
nontrivial solutions $(k^2\ne 0)$,
satisfying a polynomial of degree $(P-1)$ in $k^2$. Thus for example,
(\ref{tf}) reduces to
\ba
k^2=\frac{2\kappa_{(1)}}{\kappa_{(2)}\,(d-2)\,(d-3)}
\ea
Note that for (\ref{int}) one has the large ratio
$(\kappa_{(1)}/\kappa_{(2)})$.

\noindent
(2) \underline {Schwarzschild-type bulk black hole in the absence of GB
terms $(P=1)$}:

In (\ref{lbr}) we set
\ba
L= b\, r^2 \nn
\ea
But for $P=1$, when ${G_{(1)}}^a_b$ are given by (\ref{gcs}) one can set
\ba
\frac{L}{r^2}=\frac{c}{r^{d-1}}+b\label{lor}
\ea
the term $c r^{-(d-1)}$ is annihilated by the factor
\ba
\op{(d-1)}\label{tst}
\ea
present in each ${G_{(1)}}^a_a$ acting on $\mors$ and hence on
$\frac{L}{r^2}$. The operator (\ref{tst}) is present for all $p$. But it
acts on $\mors^p$ and $\mors^{(p-1)}$ and hence $c$ is not eliminated.

In \cite{Chakrabarti:2001di} a polynomial equation was solved for
$\frac{L}{r^2}$ in the presence of GB terms (as we do for $k^2$ here).
But the presence of $y$ and $f(y)$ no longer permit that.

Since $c$ in (\ref{lor}) is eliminated (for $P=1$) to start with, in the
equations of motion one can proceed as before with (\ref{lor}) replacing
(\ref{lbr}) and maintaining (\ref{fcc}),(\ref{fccs}), (\ref{cc}) and
(\ref{kl}). {\em One thus obtains a Schwarzschild type black hole in the
nonfactorizable case} in the absence of GB terms.) The lapse fraction
$f(y)(1-L(r))$ now depends on $y$. But the horizon ($L(r)=1$) still
depends only on $r$, and neither on $t$ nor on $y$.

{\setcounter{equation}{0}}
\section{Insertion of branes}
Having presented our bulk solutions, we now consider the modifications
necessary for inserting two positive tension branes in a fashion quite
analogous to the treatment of \cite{Kogan:2000vb}. In
the notation of \cite{Kogan:2000vb} our $f(y)$ is $a(z)^2$ where
\ba
a(y)=\frac{\cosh(k\,(y_0-|y|)}{\cosh (k\,y_0)}=\cosh k|y|-(\tanh) ky_0
\ \sinh k|y|\ .\label{4.1}
\ea
We now set, introducing orbifold symmetry permitting branes,
\ba
f(y)=a^2(y)= \frac{2+e^{-2k y_0+2 k|y|}+e^{2 k y_0-2 k|y|}}{\left(e^{k y_0}+
e^{-k y_0}\right)^2}
\label{ftwo}
\ea
Comparing to (\ref{fcc}) (along with $|y|$ for $y$) one has
\ba
\left\{c_0,c_1,c_2\right\}=\left(e^{k y_0}+e^{-k y_0}\right)^{-2}\,
\left\{2,e^{-2k y_0},e^{2 k y_0}\right\}
\ea
satisfying (\ref{cc}).\footnote{Note that for $b=0$, i.e. $L=0$, one can
only have $a(y)\sim e^{\mp k |y|}$.} Hence for $y>0$, in the bulk, our
previous solution holds with (\ref{ftwo}) and
\ba
1-L(r)=1+\frac{4 k^2}{\left(e^{k y}+e^{-k y_0}\right)^2}\,r^2
\ea
where $k^2$ is determined from (\ref{tel}).

But now one must match coefficients of $\delta$-functions in derivatives
of $a(y)$ to those contributed by the brane tensions.

Let us suppose the two branes are inserted respectively at
\[
y=0 \quad {\rm and}\quad y=Y.
\]

Note, that, from (\ref{hone}) and (\ref{htwo}), using (\ref{ftwo}),
\ba
h_1=\left(a^{-1}a_y\right)^2\ ,\quad h_2=a^{-1}a_{yy}\ ,\quad
\left(a_y=\frac{da}{dy}\right)\ .
\ea
Singular terms will arise from $h_2$ only which appears
{\it linearly} in all ${G_{(p)}}^a_a (a=t,r,i)$ and not in
${G_{(b)}}^y_y$. One obtains, in a fashion analogous to
\cite{Kogan:2000vb},
\ba
a_{yy}=k^2\,a+\lambda_1\,\delta(y)+\lambda_1\,\delta(y-Y)\label{4.6}
\ea
where
\[
\lambda_1=2a_y(0^+)\quad ,\quad \lambda_2=-2a_y(Y^-)\ .
\]
Hence
\ba
h_2&=&k^2+\frac{2a_y(0^+)}{a(0)}\delta(y)-\frac{2a_y(Y^-)}{a(Y)}
\delta(y-Y)\label{4.7}\\ &\equiv&k^2+\hat h_2\ \nn,
\ea
$\hat h_2$ denoting the singular part.

In (\ref{gptt}), (\ref{gpii}) and (\ref{gpyy}) we now set, as in
(\ref{tn})
\[
\frac{M}{r^2}=-k^2
\]
but $h_2=k^2+\hat h_2$ has now an additional singular part as in
(\ref{4.7}). Thus one obtains, with (\ref{tten}) denoting ``bulk'' values,
\ba
{G_{(p)}}^a_a&=&{\left({G_{(p)}}^a_a\right)}_{bulk}
+\frac{(-1)^{p-1}}{2^{p-1}(p-1)!}
(d-1)(d-2)\dots(d-2p+1) (k^2)^{(p-1)}\hat{h}_2\ , \nn\\
&~&\hspace{8cm}(a=t,r,i_1,\dots,i_{d-2})\label{4.8}\\
{G_{(p)}}^y_y&=&{\left({G_{(p)}}^y_y\right)}_{bulk}\ .\nn
\ea
The bulk terms contribute to match the bulk $\Lambda\ \delta^a_b$ as in
(\ref{3.4}). But now (\ref{3.4}) is modified to match the brane
tensions ($\Lambda_1$ and $\Lambda_2$, say, for the branes at $y=0$ and
$y=Y$ respectively) by the contributions induced by the terms
proportional to $\hat h_2$ in (\ref{4.8}). This gives
\ba
&~&\left\{\sum_{p=1}^P\kappa_{(p)}
\frac{(-1)^{p-1}}{2^{p-1}\,(p-1)!}\,(d-1)\dots(d-2p+1) (k^2)^{p-1}\right\}
(a(y))^{-1}\left(\lambda_1\,\delta(y)+\lambda_2\,\delta(y-Y)\right)
\nn\\&~&\hspace{9cm} \ =\Lambda_1\,\delta(y)+\Lambda_2\,\delta(y-Y)
\label{4.9}
\ea
Setting,
\ba
K_{(P)}=\left\{\sum_{p=1}^P\kappa_{(p)}\frac{(-1)^{p-1}}{2^{p-1}\,(p-1)!}
\,(d-1)\dots(d-2p+1)(k^2)^{p-1}\right\}\label{4.10}
\ea
and using (4.6) and (4.7), one obtains
\ba
\Lambda_1&=&K_{(P)}\,\ 2k\,\tanh ky_0
\label{4.11}\\
\Lambda_2&=&K_{(P)}\,\ 2k\,
\frac{\sinh k(y_0-Y)}{\cosh ky_0}\ . \label{4.12}
\ea
For $y_0>Y>0$ and $\kappa_{(1)}\gg\kappa_{(p)},\ (p=2,3,...)$,
$\Lambda_1$ and $\Lambda_2$ are both positive. The results remain
formally similar to the corresponding ones of \cite{Kogan:2000vb}
for the following reason. The linearity of each ${G_{(p)}}^a_a\quad
(a=t,r,i)$ of (\ref{4.8}) in $h_2$ (and hence $\hat h_2$) assure terms
{\it linear} in the $\delta$'s consistently on both sides of (\ref{4.9}).
Furthermore the cumulative effect of all the superposed GB terms, is
accounted for by the factor $K_{(P)}$, (\ref{4.10}).

\section{Remarks}

In the previous sections we have proceeded in successive steps: a
particularly compact and convenient construction of the generalised
Einstein tensors, then the construction of bulk solutions and finally
the insertion of branes.

The two functions $h_2(y)$ and $M(r,y)$ introduced in (\ref{htwo}) and
(\ref{mm}) respectively, played a crucial role in our construction.
On writing the warp function $f(y)$ as
\[
f(y)\ =\ a^2(y)
\]
these functions are expressed as
\ba
h_2(y)&=&a^{-1}(y)\frac{d^2a(y)}{dy^2}\label{5.1}\\
M(r,y)&=&a^{-2}(y)\left(L(r)-r^2\left(\frac{da(y)}{dy}\right)^2\right)\ .
\label{5.2}
\ea
It was shown how simply and systematically the results of
\cite{Chakrabarti:2001di} (where the coordinate $y$ was absent and the
warp factor was unity) can be generalised to the present case by simply
replacing the role of the function $\left(\frac{L(r)}{r^2}\right)$ in
\cite{Chakrabarti:2001di} with the function
$\left(\frac{M(r)}{r^2}\right)$ here, as well as adding a term linear in
$h_2(y)$ in $G_t^t,\ G_r^r,\ G_i^i\ \ (i=1,2,...,d-2)$. There is now,
of course, an additional component of the Einstein tensor $G_y^y$ which
has simply depends only on $\left(\frac{M(r)}{r^2}\right)$ but not on
$h_2(y)$. (See Eqns. (\ref{2.27}), (\ref{2.28}) and (\ref{2.29}).) Our
formalism directly selects out these crucial functions.

One can now envisage the following steps. Linear perturbations of the
metric will reveal the consequences, in our context, of the GB terms on
the zero and Kaluza-Klein modes. In particular, how the relevant results
of \cite{Kogan:2000vb,loc} are generalised.

Also, rather than seeking possible localisation of gravity in
$d$-dimensions, one can introduce intersecting branes~\cite{K},
leading to suitably lower dimensional intersections. These aspects will
be studied elsewhere.



\begin{thebibliography}{99}
\bibitem{Chakrabarti:2001di}
A.~Chakrabarti and D.~H.~Tchrakian,
``Gravitation with superposed Gauss-Bonnet terms in higher dimensions:  Black hole metrics and maximal extensions,''
Phys.\ Rev.\ D {\bf 65} (2002) 024029
[arXiv:hep-th/0101160].

\bibitem{effstr}
B.~Zwiebach,
``Curvature Squared Terms And String Theories,''
Phys.\ Lett.\ B {\bf 156} (1985) 315;
\\
R.~R.~Metsaev and A.~A.~Tseytlin,
``Order Alpha-Prime (Two Loop) Equivalence Of The String Equations Of Motion And The Sigma Model Weyl Invariance Conditions: Dependence On The
Dilaton And The Antisymmetric Tensor,''
Nucl.\ Phys.\ B {\bf 293} (1987) 385.

\bibitem{nf}
V.~A.~Rubakov and M.~E.~Shaposhnikov,
``Extra Space-Time Dimensions: Towards A Solution To The Cosmological Constant Problem,''
Phys.\ Lett.\ B {\bf 125} (1983) 139;
\\
L.~Randall and R.~Sundrum,
``An alternative to compactification,"
Phys.\ Rev.\ Lett.\  {\bf 83} (1999) 4690
[arXiv:hep-th/9906064];
\\
L.~Randall and R.~Sundrum,
``A large mass hierarchy from a small extra dimension,''
Phys.\ Rev.\ Lett.\  {\bf 83} (1999) 3370
[arXiv:hep-ph/9905221].

\bibitem{Kogan:2000vb}
I.~I.~Kogan, S.~Mouslopoulos and A.~Papazoglou,
``A new bigravity model with exclusively positive branes,"
Phys.\ Lett.\ {\bf B 501} (2001) 140
[arXiv:hep-th/0011141].

\bibitem{R}
V.~A.~Rubakov, ``Large and infinite extra dimensions: An
introduction," Uspekhi Fiz. Nauk (in press) [arXiv:hep-ph/0104152]
and references therein.

\bibitem{allgb}
J.~E.~Kim, B.~Kyae and H.~M.~Lee,
``Effective Gauss-Bonnet interaction in Randall-Sundrum compactification,''
Phys.\ Rev.\ D {\bf 62} (2000) 045013
[arXiv:hep-ph/9912344];
\\
I.~Low and A,~Zee,`` Naked singularity and Gauss-Bonnet terms in brane
world scenarios," Nucl. Phys. {\bf B 585} (2000) 395-404
[arXiv:hep-th/0004124];
\\
N.~E.~Mavromatos and J.~Rizos,
``String inspired higher-curvature terms and the Randall-Sundrum  scenario,''
Phys.\ Rev.\ D {\bf 62} (2000) 124004
[arXiv:hep-th/0008074];
\\
I.~P.~Neupane,
``Consistency of higher derivative gravity in the brane background,''
JHEP {\bf 0009} (2000) 040
[arXiv:hep-th/0008190];
\\
J.~E.~Kim and H.~M.~Lee,
``Gravity in the Einstein-Gauss-Bonnet theory with the Randall-Sundrum  background,''
Nucl.\ Phys.\ B {\bf 602} (2001) 346
[Erratum-ibid.\ B {\bf 619} (2001) 763]
[arXiv:hep-th/0010093];
\\
K.~A.~Meissner and M.~Olechowski,
``Domain walls without cosmological constant in higher order gravity,''
Phys.\ Rev.\ Lett.\  {\bf 86} (2001) 3708
[arXiv:hep-th/0009122].


\bibitem{loc}
I.~P.~Neupane,
``Localized gravity with higher curvature terms,''
arXiv:hep-th/0106100;
\\
K.~A.~Meissner and M.~Olechowski,
``Brane localization of gravity in higher derivative theory,''
Phys.\ Rev.\ D {\bf 65} (2002) 064017
[arXiv:hep-th/0106203].


\bibitem{gbcosmo}
N.~Deruelle and T.~Dole\v zel, ``Brane versus shell cosmologies in
Einstein and Einstein--Gauss-Bonnet theories," Phys. Rev. {\bf D 62}
(2000) 103502 [arXiv:gr-qc/0004021];
\\
S.~Nojiri and S.~D.~Odintsov,
``Brane-world cosmology in higher derivative gravity or warped  compactification in the next-to-leading order of AdS/CFT  correspondence,''
JHEP {\bf 0007} (2000) 049
[arXiv:hep-th/0006232];
\\
B.~Abdesselam and N.~Mohammedi,
``Brane world cosmology with Gauss-Bonnet interaction,''
arXiv:hep-th/0110143;
\\
C.~Charmousis and J.~F.~Dufaux,
``General Gauss-Bonnet brane cosmology,''
[arXiv:hep-th/0202107];
\\
J.~E.~Lidsey, S.~Nojiri and S.~D.~Odintsov,
``Braneworld Cosmology in (Anti)--de Sitter Einstein--Gauss--Bonnet--Maxwell Gravity,''
arXiv:hep-th/0202198, and references therein.

\bibitem{K}
N.~Kaloper, ``Crystal manifold universes in AdS space," Phys. Lett.
{\bf B 474} (2000) 269-281 [arXiv:hep-th/9912125].

\end{thebibliography}
\end{document}